\documentclass[conference]{IEEEtran}

\IEEEoverridecommandlockouts
\usepackage{amsmath,amssymb,amsfonts}
\usepackage{algorithmic}
\usepackage{graphicx}
\usepackage{textcomp}
\usepackage{xcolor}
\usepackage[authoryear]{natbib}
\usepackage{hyperref}
\usepackage{amssymb}
\usepackage{marvosym}
\usepackage{bm}  % 加粗包
\def\BibTeX{{\rm B\kern-.05em{\sc i\kern-.025em b}\kern-.08em
    T\kern-.1667em\lower.7ex\hbox{E}\kern-.125emX}}
\begin{document}

\title{InternVQA: Advancing Compressed Video Quality Assessment with Distilling Large Foundation Model\\

\author{%
  Fengbin Guan, Zihao Yu, Yiting Lu, Xin Li\textsuperscript{~\Letter}, and Zhibo Chen\textsuperscript{~\Letter} \\ 
  University of Science and Technology of China\\
  \texttt{\small{\{guanfb,yuzihao,luyt31415\}@mail.ustc.edu.cn, \{xin.li,chenzhibo\}@ustc.edu.cn}} \\
  }

}

% \and
% \and
% \and
% \and
% \and

\maketitle
\renewcommand{\thefootnote}{}
\footnotetext{$^{\textrm{\Letter}}$  Corresponding authors. This work was supported in part by NSFC under Grant 62371434, 62021001.}
% \footnotemark\footnotetext{$^{\textrm{\Letter}}$  Corresponding authors. This work was supported in part by NSFC under Grant 62371434, 62021001.}
  
\begin{abstract}
Video quality assessment tasks rely heavily on the rich features required for video understanding, such as semantic information, texture, and temporal motion. The existing video foundational model, InternVideo2, has demonstrated strong potential in video understanding tasks due to its large parameter size and large-scale multimodal data pertaining. Building on this, we explored the transferability of InternVideo2 to video quality assessment under compression scenarios. To design a lightweight model suitable for this task, we proposed a distillation method to equip the smaller model with rich compression quality priors. Additionally, we examined the performance of different backbones during the distillation process. The results showed that, compared to other methods, our lightweight model distilled from InternVideo2 achieved excellent performance in compression video quality assessment.

\end{abstract}


\section{Introduction}
In video understanding tasks, critical information such as semantic content, texture details, and temporal motion features is equally essential for video quality assessment(VQA). For example, human perception of distortions is significantly influenced by semantic information\citep{Qmamba}, motion blur is closely tied to the temporal motion in videos\citep{discovqa,yuan2023capturing}, and many distortions are reflected in the texture details\citep{MDVQA}. Therefore, we can leverage the strong representational power of foundational video models to enhance video quality assessment tasks.

InternVideo2\citep{internvideo2}, a new family of video foundation models, has demonstrated exceptional capabilities across various video understanding tasks, including action recognition/detection, video-text tasks, and video-centric dialogue. This success is attributed to its massive parameter size, large-scale data support, and the application of masked video learning, which endows InternVideo2 with powerful representation abilities. These rich features effectively support various downstream video tasks. 

In this study, we explore the potential of InternVideo2 for compressed video quality assessment. However, its substantial parameter count results in excessive resource consumption for these tasks. To address this, we developed a distillation method that retains the rich compression quality priors from the large model, creating a more efficient and lightweight model. This approach improves efficiency and reduces resource requirements without sacrificing performance. Additionally, we explored distilling the robust representations of InternVideo2 into various backbones, comparing homologous and heterogeneous distillation methods. Our experimental results demonstrate that, through homologous distillation, the smaller model not only surpasses existing methods on two compression quality assessment datasets but also matches or exceeds the performance of the original large model, achieving both high efficiency and superior performance.

Our contributions are as follows:

\begin{itemize}
    \item We explored the transferability of the strong video representation capabilities of InternVideo2 to video quality assessment tasks. The experimental results indicate that these representations are well-suited for quality assessment, delivering exceptional performance.
    \item To design a lightweight model tailored for compressed video quality assessment, we proposed a distillation method that extracts knowledge from InternVideo2, enabling a small model to maintain high performance while significantly reducing model size.
    \item We investigated different backbones for distillation and found that homologous distillation outperforms heterogeneous distillation, exceeding the performance of existing video quality assessment methods and achieving an optimal balance between efficiency and performance.
\end{itemize}

\begin{figure*}
    \centering
    \includegraphics[width=0.95\linewidth]{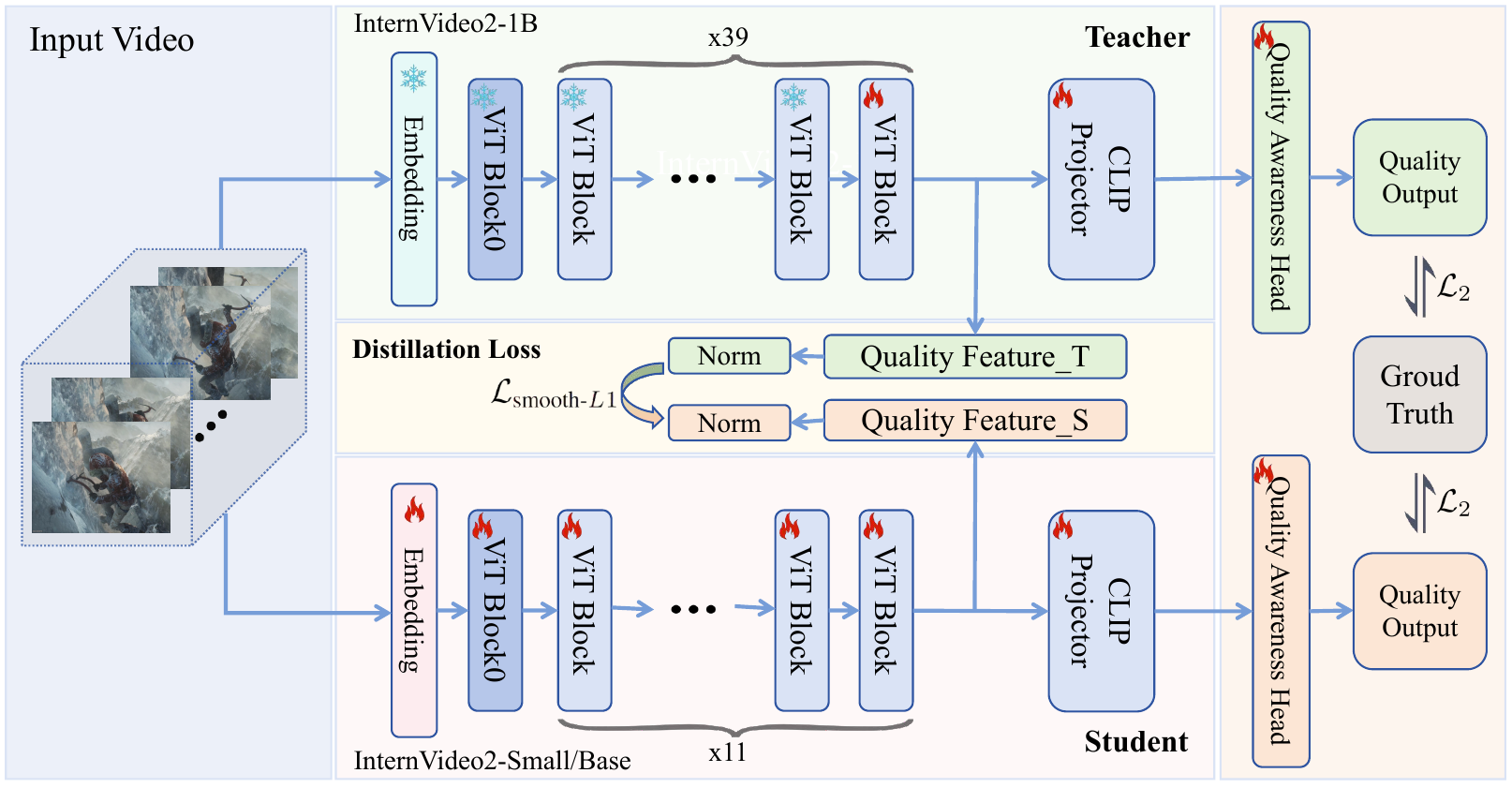}
    \caption{\normalsize Illustration of the Model Structure and Distillation Process.}
    \label{fig:  Model Structure and Distillation}
\end{figure*}

\section{Related work}

\subsection{Video Quality Assessment}
In the field of video quality assessment, there are currently two main approaches: traditional methods based on handcrafted feature extraction and modern methods based on deep learning. Traditional methods\citep{ mittal2015completely,liao2022exploring,korhonen2019two} rely on manually designed features to assess video quality; however, due to the limitations of handcrafted features, these methods typically only capture shallow quality representations and fail to account for the complex factors that influence video quality.

With the rapid advancement of deep learning, models\citep{simpleVQA,fastVQA,kvq, SwinVQA,SFIQA,AIGCVQA} based on this technology have demonstrated superior feature extraction capabilities in video quality assessment tasks. Notably, networks such as 3D-CNN\citep{3D-CNN} and Video Swin Transformers\citep{swin3D} are able to capture deep spatiotemporal features from video data. However, while these models have improved video quality assessment capabilities, they still struggle to fully capture the complex characteristics that affect video quality, making it difficult to meet the demands of comprehensive video quality assessment tasks.

\subsection{Video Foundation Models}

Video foundation models are commonly used for video understanding tasks such as action recognition, typically leveraging convolutional\citep{3D-CNN} and attention mechanisms\citep{swin3D} to extract video features. UniFormer\citep{uniformer} combines convolution and attention to reduce spatiotemporal redundancy and mitigate global dependency issues. Building on this, UniFormer2\citep{uniformerv2} incorporates pre-trained ViTs to further capture rich image priors. InternVideo\citep{internvideo} enhances the understanding of complex video content by integrating UniFormer2 with video-language contrastive learning and video-masked modeling. InternVideo2\citep{internvideo2} adopts a progressive training strategy, leveraging its massive parameter size and extensive data through masked distillation training, achieving superior performance.

\section{Method}

\subsection{Exploring the Perceptual Capabilities of Video Foundation Models}\label{AA}

InternVideo2 is selected as the video foundational model for our video quality assessment tasks due to its exceptional capability in capturing rich and intricate features required for comprehensive video understanding. As a new family of video foundation models, InternVideo2 has demonstrated strong performance across a variety of video-level tasks, including action recognition, detection, video-text matching, and video-centric dialogue. These tasks require deep semantic understanding, fine-grained texture analysis, and temporal motion tracking, all of which are crucial for both video comprehension and quality assessment.

The strength of InternVideo2 lies in its ability to process vast video data and use masked video modeling to learn highly informative representations. These representations are particularly beneficial for downstream tasks, encapsulating essential content, structure, and dynamics of the video. The model’s large parameter size, combined with extensive training on diverse datasets, provides strong generalization across various tasks and scenarios, making it ideal for quality assessment.

For video quality assessment, the perceptual impact of distortions often hinges on certain key features. Semantic information plays a critical role in human perception of video quality, as viewers are generally more sensitive to distortions in areas containing meaningful content (e.g., faces, objects) compared to less significant regions. The ability of InternVideo2 to capture semantic content ensures that such crucial areas are well-represented, enabling the model to detect and assess quality degradations that are more noticeable to human viewers.

In addition to semantic content, texture features are essential for evaluating video quality, particularly in the context of compression artifacts and other distortions. Various forms of degradation, such as blocking, ringing, or blurring, directly impact the fine details of texture. InternVideo2 excels at capturing texture information, making it ideally suited for identifying these subtle distortions and thereby contributing to more accurate quality predictions.

Furthermore, temporal motion features are essential in video understanding and play a critical role in evaluating quality distortions. Motion-related artifacts, such as motion blur or frame drops, substantially impact video quality by disrupting temporal continuity. InternVideo2 demonstrates a strong capacity to capture temporal dynamics, enabling accurate representation and detection of these motion-related distortions, which are frequently disregarded by models focused exclusively on spatial quality.

Leveraging the capacity of InternVideo2 to capture rich features—semantic, texture, and temporal motion—this approach addresses the unique challenges of video quality assessment. The powerful representations of InternVideo2 provide a strong foundation for building a robust VQA model, ensuring it generalizes across various distortions while remaining sensitive to the nuanced ways these distortions impact perceived quality.

\subsection{Knowledge Distillation Framework for Compression Video Quality Assessment}
In this study, we apply knowledge distillation\citep{knowledgedistillation} to transfer the powerful representation capabilities of the teacher model, InternVideo2-1B, to a lightweight student model to address the resource consumption challenges in compression video quality assessment. The core idea of knowledge distillation is to leverage the large teacher model to guide the learning process of the smaller student model, enabling the student to maintain efficiency while acquiring similar feature representation capabilities, especially those critical for compression quality assessment.

To ensure that the student model effectively learns the rich features related to compression quality from the teacher model, we designed a dual-loss mechanism for the distillation process. The total loss consists of three components: (1) \(\mathcal{L}_2\) Loss for the teacher model, which supervises the error between the predicted output of the teacher model and the ground truth; (2) \(\mathcal{L}_2\) Loss for the student model, which ensures that the predictions of the student model align with the ground truth; and (3) Smooth \(\mathcal{L}_1\) Loss, which aligns the feature representations between the teacher model and the student model. The overall loss function is:

\[
\mathcal{L}_{\text{total}} = \mathcal{L}_2^{\text{teacher}} + \mathcal{L}_2^{\text{student}} + \mathcal{L}_{\text{Smooth} \, \mathcal{L}_1}.
\]

This dual-loss design enables the student model to learn from both the ground truth and the feature space of the teacher model, making it more robust in handling compression distortions in video quality assessment.

Building on the design of the above distillation method, we implemented distillation in two ways:

\subsubsection{Homologous Distillation}

In homologous distillation, we selected a student model with a structure similar to that of the teacher model, maintaining architectural consistency between the two. Specifically, the teacher model is InternVideo2-1B, while the student model is the lightweight InternVideo2-Small/Base, both of which are based on the ViT (Vision Transformer) architecture. By reducing the number of parameters and layers in the student model, we effectively lower the model complexity while preserving the key feature extraction capabilities of the original architecture.

As shown in Figure \ref{fig:  Model Structure and Distillation}, in the homologous distillation process, we train only the last ViT block and head of the teacher model, freezing the rest of the model. In contrast, the entire student model is trainable. During distillation, the feature outputs of the final layer from both the teacher and student models are aligned using Smooth L1 Loss, ensuring that the student model can effectively learn the high-level representations from the teacher model. Additionally, we use ground truth to supervise the final outputs of both models to ensure prediction accuracy. Under this homologous distillation framework, the student model successfully inherits the compression quality priors from the teacher model, particularly excelling in capturing the semantic, texture, and temporal motion features essential for video quality perception.

\subsubsection{Heterogeneous Distillation}
In heterogeneous distillation, we adopted a student model with a different architecture from that of the teacher model. Specifically, the teacher model remains InternVideo2-1B, while the student model is FastVQA, with a backbone based on 3D Swin Transformer. Although FastVQA is a lightweight model that performs well in video quality assessment tasks, the significant architectural differences between it and the teacher model make the feature transfer process more complex.

Similar to homologous distillation, we use Smooth L1 Loss to align the final-layer feature outputs between the teacher and student models, ensuring that the student model inherits important compression quality priors from the teacher model. However, due to the substantial architectural differences, the student model faces greater challenges in feature learning and transfer, especially when dealing with high-frequency texture details and temporal motion features related to compression distortions.

Through these two distillation strategies, we explore the performance of homologous and heterogeneous architectures in the knowledge distillation process, analyzing how different distillation approaches influence the representation capabilities of the student model without altering the teacher model.

\begin{table*}[htbp]
\caption{Comparison of Quality Metrics on BVI-HD and Waterloo IVC 4K Datasets}
\renewcommand{\arraystretch}{1.7} % 增大行间距
\begin{center}
\begin{tabular}{l|c|cc|cc}
\hline
& & \multicolumn{2}{c|}{BVI-HD} & \multicolumn{2}{c}{Waterloo IVC 4K} \\ 
                        {Method} & {Params}                         & PLCC         & SRCC         & PLCC        & SRCC        \\ \hline
SimpleVQA\citep{simpleVQA}      & 25.93M        & 0.714            & 0.731            & 0.563           & 0.557           \\
KSVQE\citep{kvq}          & 46.16M        & 0.790            & {0.756}            & 0.740           & 0.767           \\
FastVQA\citep{fastVQA}        & 92.22M       & 0.655            & 0.623            & 0.780           & 0.734           \\
InternVQA-Small  & 23.05M       & 0.686            & 0.630            & 0.751           & 0.766           \\
InternVQA-Base  & 89.44M       & 0.752            & 0.688            & 0.751           & 0.759           \\
InternVQA-1B & 1020.71M       & {0.792}            & \textcolor{red}{0.761}            & \textcolor{red}{0.849}           & \textcolor{red}{0.891}           \\
\hline
FastVQA\textsubscript{dist} & 92.22M  & 0.714 ($\uparrow$0.059)            & 0.644 ($\uparrow$0.021)            & 0.790 ($\uparrow$0.010)           & 0.742 ($\uparrow$0.008)           \\
InternVQA\textsubscript{dist}-Small & 23.05M & 0.759 (\textcolor{blue}{$\uparrow$0.071})            & 0.734 (\textcolor{blue}{$\uparrow$0.104})            & 0.821 ($\uparrow$0.070)           & 0.794 ($\uparrow$0.028)           \\
InternVQA\textsubscript{dist}-Base & 89.44M & \textcolor{red}{0.813} ($\uparrow$0.061)            & 0.711 ($\uparrow$0.023)            & {0.824} (\textcolor{blue}{$\uparrow$0.073})           & {0.852} (\textcolor{blue}{$\uparrow$0.093})           \\ \hline
 \multicolumn{6}{l}{$^{\mathrm{*}}$\textcolor{red}{Red} indicates the best performance, and \textcolor{blue}{Blue} indicates the highest performance gain.}
\end{tabular}
\end{center}
\label{Table: Performance Comparison}
\end{table*}

\section{Experiments}
\subsection{Experimental Setup}
\subsubsection{Datasets}
In this study, we utilized two compression video quality assessment datasets: the BVI-HD dataset \citep{BVIHD} and the Waterloo IVC 4K Video Quality Database \citep{Waterloo4K}.

The BVI-HD dataset includes 32 reference videos and 384 distorted sequences, covering 12 types of distortions generated using HEVC and HEVC with synthesis mode (HEVC-SYNTH), offering a foundation for studying video quality under these compression methods.

The Waterloo IVC 4K Video Quality Database consists of 20 pristine 4K videos encoded using five different encoders (HEVC, H264, VP9, AV1, AVS2) at three resolutions (960x540, 1920x1080, 3840x2160). Each resolution has four distortion levels, yielding 1200 videos in total, providing a rich resource for evaluating compression performance across different encoders and resolutions.

These two datasets offer diverse types of compression distortions and various resolution settings, enabling a comprehensive evaluation of the video quality assessment performance of our model across different compression scenarios.

\subsubsection{Evaluation Criteria}

In this study, we use Pearson Linear Correlation Coefficient (PLCC)\citep{plcc} and Spearman's Rank Correlation Coefficient (SRCC)\citep{srcc} as evaluation metrics, both of which have a range between 0 and 1. A value nearer to 1 indicates a higher correlation between the predicted outcomes and the actual ground truth.

\subsubsection{Experimental Details}

Our experimental approach closely follows the training strategy outlined in InternVideo2\citep{internvideo2}. The input videos are randomly cropped into patches of 8×224×224 resolution. We employed InternVideo2-1B as the teacher model, which has an embedding dimension of 1408 and a depth of 40 ViT blocks. For the student models, we used InternVideo2-Small/Base, where the depth of their ViT blocks is 12. Specifically, the embedding dimensions for the Small model and Base model are 384 and 768, respectively. The training process utilized pre-trained weights from K400, with a batch size of 8. All experiments were conducted using DeepSpeed on two A100 GPUs.

\subsection{Experimental Performance Comparison and Analysis}
To evaluate the capability of InternVideo2 in compression video quality assessment, we conducted experiments on the two aforementioned datasets, with the results presented in Table\ref{Table: Performance Comparison}. InternVideo2-1B, leveraging its strong video quality representation ability, outperformed other existing VQA methods, achieving the best performance. In contrast, the performance of the InternVideo2-Small/Base models was not as competitive. To develop a lightweight and high-performance model for compression quality assessment, we applied knowledge distillation to transfer the representation capabilities of InternVideo2-1B to the smaller models.

The experimental results demonstrate that the designed distillation method significantly enhances the performance of the student models, leading to an efficient and high-performing compression quality assessment model. Specifically, under homologous distillation, the student models (InternVideo2-Small/Base), which share the same ViT architecture as the teacher model (InternVideo2-1B), were better able to inherit the feature extraction abilities of the teacher model. This was particularly evident in handling complex video distortions, where the student models performed comparably to, and in some cases even surpassed, the teacher model. The student models effectively balanced computational efficiency with performance.

In contrast, although heterogeneous distillation enabled a certain degree of model lightweighting, the structural differences between the 3D Swin backbone of FastVQA and the ViT architecture of InternVideo2-1B presented challenges for the student model to fully replicate the feature extraction capabilities of the teacher model. This limitation was particularly evident in the processing of temporal motion information and high-frequency texture details, where the performance of FastVQA was noticeably inferior to that of the student models trained through homologous distillation. The performance gap indicates that, while heterogeneous distillation offers some architectural flexibility, it proves less effective than homologous distillation for the task of compressed video quality assessment.

% \newpage

\section{Conclusion}
In this study, we demonstrated the substantial potential of InternVideo2 in the field of compression video quality assessment. By leveraging its rich video representation capabilities, we designed a distillation method to transfer knowledge from the large model to a smaller, more efficient model. Our experiments reveal that the lightweight model achieved exceptional performance, closely approaching or even surpassing the results of the original large model, especially through homologous distillation. This method effectively balances high performance with resource efficiency, offering a significant reduction in computational requirements while maintaining strong predictive accuracy. These findings underscore the effectiveness of distilling large foundational models for lightweight, high-performance compression video quality assessment.

\vspace{12pt}

\bibliographystyle{abbrvnat}
\bibliography{ref}
\end{document}